# Harnessing Layer-Controlled Two-dimensional Semiconductors for Photoelectrochemical Energy Storage via Quantum Capacitance and Band Nesting

Praveen Kumar[1], Tushar Waghmare[1,2], Sudhir Kumar[1], Rajdeep Banerjee[4], Suman Kumar Chakraborty,[1] Subrata Ghosh[3], Dipak Kumar Goswami[4], Sankha Mukherjee[2], Debabrata Pradhan[1*], Prasana Kumar Sahoo[1*]

[1]Materials Science Centre, Indian Institute of Technology Kharagpur, 721302, India.

[2]Department of Metallurgical & Materials Engineering, Indian Institute of Technology Kharagpur, 721302, West Bengal, India.

[3]Micro and Nanostructured Materials Laboratory (NanoLab), Department of Energy, Politecnico di Milano, Milano 20133, Italy.

[4]Department of Physics, Indian Institute of Technology Kharagpur, 721302, India

*prasana@matsc.iitkgp.ac.in; deb@matsc.iitkgp.ac.in; sankha@metal.iitkgp.ac.in

## ABSTRACT

Two-dimensional (2D) transition metal dichalcogenides such as molybdenum diselenide ($MoSe_2$) semiconductors have emerged as promising materials for applications in optoelectronics and energy storage due to their exceptional layer-dependent bandgap. Despite several proof-of-concept applications, the scalable and controlled synthesis of high-quality 2D $MoSe_2$ layers remains challenging. Particularly, traditional hydrothermal and liquid-phase exfoliation methods, commonly used in energy storage, lack precision in layer control and their exact role at the nanoscale, limiting further applications. Atmospheric pressure chemical vapor deposition (AP-CVD) offers a scalable solution, enabling the growth of high-quality, large-area, and layer-controlled 2D $MoSe_2$. However, the photoelectrochemical performance of such CVD-grown 2D $MoSe_2$, especially their layer-dependent energy storage capabilities, remains largely unexplored. This work addresses this gap by probing the $MoSe_2$ layer-dependent unique quantum capacitance and photo-induced charge storage properties. Employing a three-electrode setup in aqueous 0.5 M $H_2SO_4$, we demonstrate a $MoSe_2$ layer-dependent increase in areal capacitance under both dark and illumination conditions. Notably, a six-layer $MoSe_2$ film exhibits the highest areal capacitance, achieving 96 µF/cm² in the darks and 115 µF/cm² under illumination at a current density of 5 µA/cm². First-principles calculations using Density Functional Theory (DFT) and Many-Body Perturbation Theory reveal that Van Hove singularities and band nesting significantly enhance optical absorption, increasing quantum capacitance. These findings underscore the potential of APCVD-grown 2D $MoSe_2$ layers as light-responsive, high-performance electrochemical energy storage electrodes, paving the way for next-generation innovative energy storage systems.

# 1. INTRODUCTION

The rapid advancements in the Internet of Things, flexible electronics and the increasing demand for sustainable energy solutions have driven significant interest in innovative energy storage technologies[1]. Among these, ultrathin film electrodes (thicknesses less than 20 nm) with efficient electrochemical responses are in demand for applications such as on-chip integration and self-powered electronic devices[2]. Layered two-dimensional (2D) materials, including graphene and transition metal dichalcogenides (TMDs) — such as $MoS_2$ and $MoSe_2$ — have garnered particular attention due to their large surface area, tunable and layer-dependent bandgap, and high carrier mobility[3–5]. Furthermore, the ability to control the number of layers and their easy exfoliation and on-demand transfer to arbitrary substrates facilitates the fabrication of wide varieties of heterostructures, which enhances critical properties such as quantum capacitance, light-matter interaction, absorption and charge-storage kinetics[6]. For example, the quantum capacitance of electrodes increases with the number of graphene layers[7]. However, graphene, despite its high surface area of 2630 cm²/g, has a limited areal (gravimetric) capacitance of 13.5 µF/cm² (equivalent to 335 F/g) due to its primary reliance on double-layer charge storage mechanisms[8]. In contrast, TMDs exhibit a synergistic combination of efficient ion intercalation, deintercalation processes, high surface area and multiple oxidation states (e.g., +2 to +4), enabling efficient charge storage through Faradaic ion intercalation into the interlayer space, in addition to electric double-layer (EDL) capacitance[9]. This has positioned TMDs as promising materials for achieving higher specific capacitance and energy density in energy storage devices[10,11].

Recently, there has been growing interest in photo-induced electrochemical energy storage devices, where light illumination enhances charge-storage performance. For instance, the photoelectrochemical properties of mono- to few-layered 2D $MoS_2$ improve electron transfer kinetics and electric double-layer capacitance with increasing number of layers, suitable for supercapacitor applications[12]. Similarly, light-enhanced supercapacitors based on 2D $ReS_2$ have demonstrated rapid charge transfer processes and high specific capacitance, making them promising materials for next-generation smart energy storage applications[13]. Compared to $MoS_2$, 2D $MoSe_2$ exhibits significant spin-orbit interaction, strong light-matter interaction characteristics, photoluminescence quantum yield, and relatively large interlayer spacings, which could facilitate electrolyte ion intercalation and associated charge transfer process. Additionally, its near-infrared band gap and superior optical absorption suggest a strong potential for light-induced energy storage applications, making 2D $MoSe_2$ an attractive material in this emerging field.[14–16] In contrast to $MoS_2$, the energy storage applications of 2D $MoSe_2$ remain relatively underexplored. Additionally, there is a growing need to investigate the influence of the layer number on photo-induced charge transfer and storage processes, which can be effectively studied using high crystalline quality and layer-controlled 2D TMDs.

A critical challenge lies in synthesizing scalable, high-quality, and large-area 2D MoSe$_2$ with controlled layer numbers suitable for practical applications. While hydrothermal synthesis has been widely used to produce MoSe$_2$ flakes, this method often leads to inconsistent thickness, poor crystalline quality and phase purity, and limited control over layer uniformity[17,18]. In contrast, the atmospheric pressure chemical vapor deposition (APCVD) technique has been shown to produce wafer scale and high-quality 2D TMDs with controlled layer number and defect density, which could offer significant advantages in scalability towards developing practical applications[19–21]. While CVD-grown 2D TMDs have been extensively studied for electrical, optical, and catalytic applications[22–25], an accurate understanding of the photo-induced electrochemical charge storage properties of 2D MoSe$_2$ can pave the way for future scalability and advanced integration.

In this work, we demonstrate the synthesis of high crystalline quality and large-area (~ 1 cm$^2$) 2D MoSe$_2$ with varying layer thicknesses using the APCVD method. We investigate their layer-dependent photoelectrochemical areal capacitance in an aqueous 0.5 M H$_2$SO$_4$ electrolyte using a three-electrode configuration. To elucidate the experimental results, we employed numerical calculations using first-principles methods, including Density Functional Theory (DFT) and the Many-Body Perturbation Theory (MBPT). Furthermore, theoretical calculations of the absorption spectra and quasi-particle band structures reveal the roles of band nesting and Van Hove singularities in enhancing optical absorption and quantum capacitance.

## 2. EXPERIMENTAL METHODS

### 2.1. Controlled growth of 2D MoSe$_2$ using APCVD

Prior to growing MoSe$_2$, the SiO$_2$/Si substrate (oxide layer thickness of 300 nm) was ultrasonicated for 10 min acetone, isopropanol, and deionized water in turn and then dried by blowing N$_2$ gas. A proportion of 300 mg of selenium (Se) powder (99.5% Alfa Aesar) and 5 mg of molybdenum oxide (MoO$_3$) powder (99.5% Alfa Aesar) were used as precursors and placed in two different zones in the quartz tube of 1-inch diameters within a CVD furnace. Se powder was placed in alumina boats at a temperature of 200 °C in Zone 1, and MoO$_3$ powder on SiO$_2$/Si was placed at 850 °C in Zone 2. The growth was carried out in the presence of an ultrahigh pure N$_2$ atmosphere with a flow rate of 200 sccm for 15 minutes, and the deposition pressure of 1 atm was maintained during the growth. Upon completion of growth, the sample was cooled down to room temperature using natural cooling and taken out for further characterization. To control the number of layers during the growth of MoSe$_2$, the oxide precursor amount used was 5 mg, 10 mg, 15 mg, and 20 mg for single-layer (1L), bilayer (2L), trilayer (3L), and multi-layer (ML), respectively.

### 2.2. Materials Characterization.

Optical imaging of the samples was performed using an Olympus optical microscope. The thicknesses of the layer-dependent MoSe$_2$ thin films were measured with an Asylum MFP-3D atomic force microscope (AFM). Raman and photoluminescence (PL) spectroscopy were conducted using an HR-Evolution Raman spectrometer (Horiba) equipped with a 532 nm laser and a 100× objective. The crystal structure of the as-grown 2D $MoSe_2$ layers was determined by X-ray diffraction (XRD) using a Malvern Panalytical Empyrean diffractometer operated at 40 mA and 45 kV during 2θ scans. X-ray photoelectron spectroscopy (XPS) measurements were carried out using a PHI 500 VersaProbe II system with a monochromatic Al Kα X-ray source (1486.6 eV) to analyze the composition and chemical states of the elements in the material. The usual PMMA-assisted wet transfer of the as-grown large-area TMD to the TEM grid was used to prepare the high-angle annular dark field scanning transmission electron microscope (HAADF-STEM) sample. A JEOL JEM-ARM300F2 microscope with a cold-field emission gun operating at 300 kV was employed for HAADF-STEM imaging. The microscope was aberration-corrected. The JEOL HAADF detector was employed to acquire the HAADF-STEM images, with the following experimental parameters: a scan speed of 20 μs per pixel and a probe size of 8c.

### 2.3. Electrode Fabrication.

To fabricate the electrode, we first spin-coated a thin layer of PMMA with 500 rpm for 1 minute onto a $MoSe_2$ thin film placed on a $SiO_2$/Si substrate, which was then dried overnight. The $MoSe_2$/PMMA film was then immersed in a KOH solution for 10 minutes to etch the $SiO_2$ layer, which released the 2D film. The $MoSe_2$/PMMA film was fished and transferred onto a deionized water container, followed by multiple washes to remove any KOH residue. Finally, the PMMA film containing $MoSe_2$ thin film is transferred to the target substrate (FTO), dried over, and dissolved in an acetone solution to remove the PMMA layer from the $MoSe_2$ thin film (Figure S3). Subsequently, the $MoSe_2$ thin film was dried at 80 °C on a hot plate for 30 minutes to improve the adhesion of the $MoSe_2$ thin film to the target FTO substrate (Figure S4).

### 2.4. Electrochemical Measurements.

Cyclic voltammetry (CV), galvanostatic charge-discharge (GCD), and electrochemical impedance spectroscopy (EIS) were carried out in 0.5 M aqueous $H_2SO_4$ solution with a CHI 760D electrochemical workstation (CH Instruments, Inc., USA). A standard three-electrode configuration was used to carry out all of the measurements, where the fabricated electrode was used as working electrodes, the saturated calomel electrode (SCE) was used as a reference, and the platinum wire was used as the counter electrode. The impedance spectroscopy measurement was carried out with 10 mV *a.c.* Perturbation within the frequency range of 10 mHz to 100 kHz. An 80 W bulb was used as the light source for studying the light-induced charge-storage performance for as-fabricated electrodes (Figure S5).

The CV measurement was performed at different scan rates (10 mV/s to 100 mV/s) in the potential range of 0 to 0.8 V. The areal capacitance was determined using the CV curves and the following equation[26]:

$$C_A = \frac{\int I \times dV}{A \times v \times \Delta V} \quad (1)$$

where C_A is the areal capacitance (µF.cm⁻²), $A$ is the electrode's surface area (cm²), $I$ is the applied current (mA), $\Delta V$ is the potential window (mV), and $v$ is the scan rate (mV.s⁻¹). The $\int IdV$ refers to the area enclosed by the corresponding CV response. Galvanostatic charge-discharge (GCD) testing was conducted at various current densities, ranging from 5 to 20 µA/cm² within a potential window of 0 to 0.8 V. Areal capacitance was calculated from the GCD curves using the following equation[26]:

$$C_A = \frac{I \times \Delta t}{A \times \Delta V} \tag{2}$$

where $C_A$ is the areal capacitance (µF.cm⁻²), A is the electrode's surface area (cm²), $\Delta t$ is the discharging time (s), $I$ is the applied current (mA), and $\Delta V$ is the potential window (mV). This method provides an accurate assessment of capacitance across various current densities.

## 3. THEORY

### 3.1 Quantum capacitance

The interfacial capacitance (IC) depends on quantum capacitance (C_Q) and electronic-double layer capacitance (EDLC) and is given by,

$$\frac{1}{IC} = \frac{1}{C_Q} + \frac{1}{EDLC} \tag{3}$$

The quantum capacitance has more effect on interfacial capacitance in the local potential window of 0 to 0.8 eV [27–29]. The quantum capacitance is defined as: $C_Q = \frac{dQ}{dV_a}$, where $dQ$ is differential charge density and $V_a$ is the MoSe₂ electrode potential. The excess charge density $Q$, is defined as:[30]

$$Q = e \int_{-\infty}^{+\infty} D(E)[f(E) - f(E - eV_a)]dE \tag{4}$$

where, $D(E)$ is the density of states and $f(E)$ represents the Fermi-Dirac distribution. $E$ represents energy relative to the fermi energy, and e is the elementary electronic charge. Therefore $C_Q$ is[30–32],

$$C_Q = e^2 \int_{-\infty}^{+\infty} \frac{D(E)}{4K_BT} sech^2\left(\frac{E - eV_a}{2K_BT}\right) dE \tag{5}$$

Since the function $sech^2(E)$ has a range of (0,1] (as shown in figure S7 in the supplementary material) and reaches its maximum at $E = 0$, the density of states near the Fermi energy can play a significant role in determining quantum capacitance. Therefore, when comparing the quantum capacitance of two different samples, it is essential to consider the density of states near the Fermi energy.

### 3.2 DFT calculations

The density of states and nudged elastic band calculations were performed using DFT with the Quantum Espresso package[33–35]. At this level, the exchange-correlation effects were described by the gradient-corrected Perdew−Burke−Ernzerhof (PBE) functional[36], while norm-conserving pseudopotentials[37] were used to model electron-ion interaction. Wavefunctions were expanded in plane waves with an energy cut-off of 50 Ry. The SCF convergence criteria for electron density was set at $1 \times 10^{-10}$ Ry/Bohr. We considered the standard 1×1-unit cell of MoSe$_2$, comprising *3n* atoms, where *n* refers to the number of layers in the structure. The Brillouin zone was sampled with a 13×13×1 Monkhorst-Pack mesh for all structures with the energy and force convergence threshold of $1 \times 10^{-6}$ and $1 \times 10^{-5}$ Ry/Bohr respectively. The van der Waals interactions were treated with the Grimme-D2 semi-empirical approach[38], and a vacuum region of 20 Å was added along the direction orthogonal to the atomic layers to ensure decoupling between periodic structures.

### 3.3 G$_0$W$_0$ calculations

The GGA functionals used to describe the exchange-correlation in the Kohn-Sham equation smooth out the derivative discontinuity when adding or removing an electron in a given system, leading to the underestimation of band gaps in DFT calculations. This underestimation affects the density of states near the Fermi energy, resulting in a relatively poor estimation of quantum capacitance. We can improve the band gap accuracy using hybrid functionals and MBPT methods. However, using MBPT provides access to the excitonic eigenvalues and wavefunctions, which is helpful in the determination of absorption spectra. In this study, we employed MBPT techniques to obtain relativistic band gap and density of states (DOS) values. The many-body correction parameters were calculated using the non-self-consistent perturbative G$_0$W$_0$ approximation, as implemented in the YAMBO code[39–41].

The exchange component of the self-energy, often referred to as the Fock term in Hartree-Fock self-energy, represents the matrix elements of the nonlocal exchange potential. It was calculated as follows:

$$\Sigma_{nk}^{x} = -\frac{1}{N_q V} \sum_{0,q} \sum_{G} |\rho_{no}(\boldsymbol{k}, \boldsymbol{q}, \boldsymbol{G})|^2 \bar{v}_G(\boldsymbol{q}) \tag{6}$$

where $V$ is the volume of a unit cell in real space, $N_q$ is the number of points in the q-grid, and $o$ represents the occupied bands. The matrix elements $\rho_{no}$ represent the overlap between Bloch states $\{n, o\}$ under reciprocal lattice translations. The $\bar{v}_G(q)$ is the average coulomb interactions in a Brillion Zone (BZ) around $q$ calculated through the Random Integration Method[42].

On the other hand, the dynamical correlation part of self-energy accounts for the many-body interactions beyond the Hartree-Fock approximation, specifically the dynamic effects of electron correlation, and it was calculated as:

$$\Sigma_{nk}^{C}(\omega) = \frac{1}{N_q V} \sum_{G,G',q} M_{GG'}^{nk}(\boldsymbol{q}, \omega) \bar{W}_{GG'}^{C} \tag{7}$$

where, the $M_{GG'}^{nk}(q, \omega)$ represents the matrix elements that describe the coupling between the electronic states through screened interactions. The $\bar{W}_{GG'}^{C}$ term corresponds to the average correlation part of the screened potential. Within the plasmon pole approximation (with Godby Needs formulation[43]), the $\overline{W}^C_{GG'}$ was written as,

$$\overline{W}^C_{GG'} = \frac{1}{D_\Gamma} \int_{D_\Gamma} \frac{2R_{GG'}(\boldsymbol{q}+\boldsymbol{q}')V_{GG'}(\boldsymbol{q}+\boldsymbol{q}')}{\omega^2 - [V_{GG'}(\boldsymbol{q}+\boldsymbol{q}') - i\eta]^2} d\boldsymbol{q}' \tag{8}$$

This approach, known as the W-averaged (w-av) method[44,45], where the correlation self-energy is written in terms of the $\overline{W}^C_{GG'}$ instead of $W^C_{GG'}$ mitigates the problem of sharp $q$ dependence in $W^C_{GG'}$ for small $G$ vectors. The integrals in equation (8) are computed using a 2D Monte Carlo integration method, while an interpolation scheme is employed to calculate $\overline{W}^C_{GG'}$. This method also accelerates the convergence of GW results with respect to k-point sampling, giving us a far reliable band gap on a moderately dense k-grid. Convergence of the polarization function with respect to G-vectors and bands, as well as correlation self-energy with bands in the correlation kernel, are presented in Figure S8 in the supplementary material.

### 3.4 Bethe-Salpeter Equation calculations

To capture excitonic effects in optical spectra, the Bethe-Salpeter Equation (BSE) was used, where the electron-hole (e–h) Green's function, $L$, described the interaction between an excited electron and a hole. The BSE is essential for accurately modeling absorption spectra, especially for 2D TMDs, given their strong excitonic binding energies. The response function, $\chi$, which is crucial for computing the macroscopic dielectric function $\epsilon_M(\omega)$ and polarizability $\alpha(\omega)$, was calculated earlier in G$_0$W$_0$ calculation using random phase approximation. The modified non-interacting response function for $\boldsymbol{q} = 0$ can be written in terms of non-interacting electron-hole Green's function, $L^0$, as:

$$\lim_{\boldsymbol{q} \to 0} \chi^0_{GG'}(\boldsymbol{q},\omega) = -i \sum_{vck} \left[ \lim_{\boldsymbol{q} \to 0} \rho^*_{cvk}(\boldsymbol{q},\boldsymbol{G}) \rho_{cvk}(\boldsymbol{q},\boldsymbol{G}') \right] L^0_{vck}(\omega) \tag{9}$$

The Bethe-Salpeter equation itself is expressed as:

$$L = L_0 + L_0 \Xi L \tag{10}$$

where $L_0$ is the non-interacting e–h Green's function, and $\Xi$ is the BSE kernel. The matrix $[\Xi = 2\overline{V}_{vck,v'k'c'} - W_{vck,v'k'c'}]$ represents the BSE kernel, where $W$ gives the screened electron-electron interaction and an exchange term $\overline{V}$ accounts for bare Coulomb interaction. The equation was simplified by reformulating it as an eigenvalue problem using the effective BSE Hamiltonian:

$$H_{vck,v'c'k'} = (\varepsilon_{ck} - \varepsilon_{vk})\delta_{cc'}\delta_{vv'}\delta_{kk'} + (f_{ck} - f_{vk})(2\overline{V}_{vck,v'k'c'} - W_{vck,v'k'c'}) \tag{11}$$

where $\varepsilon_{nk}$ are single-particle energies and $f_{ck}$ are occupation factors. The $(\varepsilon_{ck} - \varepsilon_{vk})$ factor is the quasiparticle corrections from the previous G$_0$W$_0$ calculations. The Hamiltonian was solved using standard Tamm-Dancoff approximation to obtain the transition energies and oscillator strengths needed for the optical spectrum. Thus, the macroscopic dielectric function $\epsilon_M(\omega)$ can be represented in terms of eigenvalues $E_\lambda$ and eigenstates $|\lambda\rangle$ as:

$$\epsilon_M(\omega) = 1 - \frac{8\pi}{|q|^2 N_q V} \sum_{vc\mathbf{k}} \sum_{v'c'\mathbf{k'}} \left[\lim_{q\to 0} \rho^*_{cv\mathbf{k}}(\mathbf{q},\mathbf{G}) \rho_{c'v'\mathbf{k'}}(\mathbf{q},\mathbf{G'})\right] \sum_\lambda \frac{A^\lambda_{cv\mathbf{k}}\left(A^\lambda_{c'v'\mathbf{k'}}\right)^*}{\omega - E_\lambda} \quad (12)$$

with $A^\lambda_{cv\mathbf{k}} = \langle vc\mathbf{k}|\lambda\rangle$ as an eigenvector of the BSE Hamiltonian. The absorption spectra were obtained by plotting the imaginary component of the macroscopic dielectric function as a function of photon energy. Convergence tests for the BSE kernel components, including the exchange contribution (BSENGexx), screened Coulomb potential matrix W (BSENGBlk), and e-h basis size dependence, are detailed in Figure S8 in the supplementary material.

## 4. RESULTS AND DISCUSSION

### 4.1. Structural Characterization

The atomic model of 1L and 2L MoSe$_2$ is depicted in Figure 1(a, b), and the corresponding optical micrograph images of the as-grown 1L and 2L MoSe$_2$ on a Si/SiO$_2$ substrate over a 1 cm$^2$ area are shown in Figure 1(c, d). The color of the as-deposited MoSe$_2$ layer on the SiO$_2$/Si substrate changes from pink to deep blue as the layer number increases by increasing the amount of transition metal oxide precursors during the CVD growth process. The thickness of as-grown films was measured using atomic force microscopy (AFM) [Figure S1(a, b, d and f)]. From the height profiles, the estimated thickness for the 2D film grown under the oxide precursor amount of 5, 10, 15, and 20 mg is 0.95 nm (1L-MoSe$_2$), 1.7 nm (2L-MoSe$_2$), 2.2 nm (3L-MoSe$_2$), and 4.4 nm (ML-MoSe$_2$), respectively. In our case, the film with a thickness of 0.95 nm is slightly higher than the thickness of ideal 1L-MoSe$_2$ (0.645), which might resulted from substrate and absorbates[46]. Figure 1(e, f) shows the HAADF-STEM images of 1L and 2L MoSe$_2$, respectively, showing the high crystalline quality of atoms with a *2H$_c$* crystal structure. In addition, the chemical composition and oxidation state of the elements were investigated using X-ray photoelectron spectroscopy. The *+4* oxidation state of molybdenum was confirmed by the presence of Mo 3$d_{3/2}$ and Mo 3$d_{5/2}$ peaks at binding energies of 232.02 and 228.75 eV, respectively [Figure 1(g)][47]. Furthermore, the 3$d$ peaks of selenium were divided into two well-defined peaks such as 3$d_{3/2}$ and 3$d_{5/2}$ [Figure 1(h)]. These peaks were observed at binding energies of 55.12 and 54.37 eV, respectively, verifying the *-2* oxidation state of selenium in MoSe$_2$ 2D layers[48].

Raman spectroscopy is a well-established, non-destructive technique for identifying the number of layers of 2D materials. The Raman spectra of layer-dependent MoSe$_2$ 2D films consist of two prominent peaks [Figure 1(i)]: the out-of-plane *A$_{1g}$* mode within the range of 240 to 243 cm$^{-1}$ and the in-plane *E$^1_{2g}$* mode in the spectral range of 284 to 287 cm$^{-1}$. The *A$_{1g}$* mode at 240 cm$^{-1}$ indicates the monolayer nature of MoSe$_2$, and it shifts to 243 cm$^{-1}$ with increasing film thickness, indicating an increase in the number of layers. Similarly, as the number of layers increases, the *E$^1_{2g}$* mode broadens, and its position redshifts from 287 cm$^{-1}$ (1L) to 284 cm$^{-1}$ with an increase in the number of layers, consistent with previous reports. The PL spectrum from individual 1L, 2L, 3L, and ML MoSe$_2$ 2D films exhibits a single prominent asymmetric emission peak [Figure 1(j)]. Notably, the PL intensity of 1L-MoSe$_2$ is 2.6, 3.3, and 4.3 times higher than that of 2L, 3L, and ML-MoSe$_2$, respectively. This decrease in PL intensity with increasing layer number is attributed to enhanced interlayer coupling and a transition from direct to indirect bandgap in the thicker 2D TMDs. The stronger interlayer coupling distorts the band structure, reducing the degree of band nesting responsible for additional optical absorption. Consequently, intensity bleaching in the PL spectra as the number of layers increases from 1L to ML is expected to improve the carrier generation, transfer and their interaction with the neighboring elements. Additionally, the emission peak position redshifts from 813 nm (1.52 eV) for 1L-MoSe$_2$ to 816 nm (1.51 eV) for ML-MoSe$_2$, primarily due to increased dielectric screening and the quasi-indirect nature of multilayer films[49]. Furthermore, the crystal structure of the as-grown ML- MoSe$_2$ (6 layers) film was probed using X-ray diffraction to study [Figure S2(a)][50], demonstrating the large-scale structural, phase-purity and quality of our as-grown 2D MoSe$_2$. The XRD diffraction peaks at 2θ values of 13.67, 27.76, 32.88, 42.04, 57.24, 61.67, 65.90, and 69.52° correspond to the (002), (004), (002), (006), (008), (220), (203), and (400) diffraction planes of hexagonal MoSe$_2$ with a *D6 4h* (*P63/mmc*) space group (JCPDS: 29-0914), respectively (Figure S2).

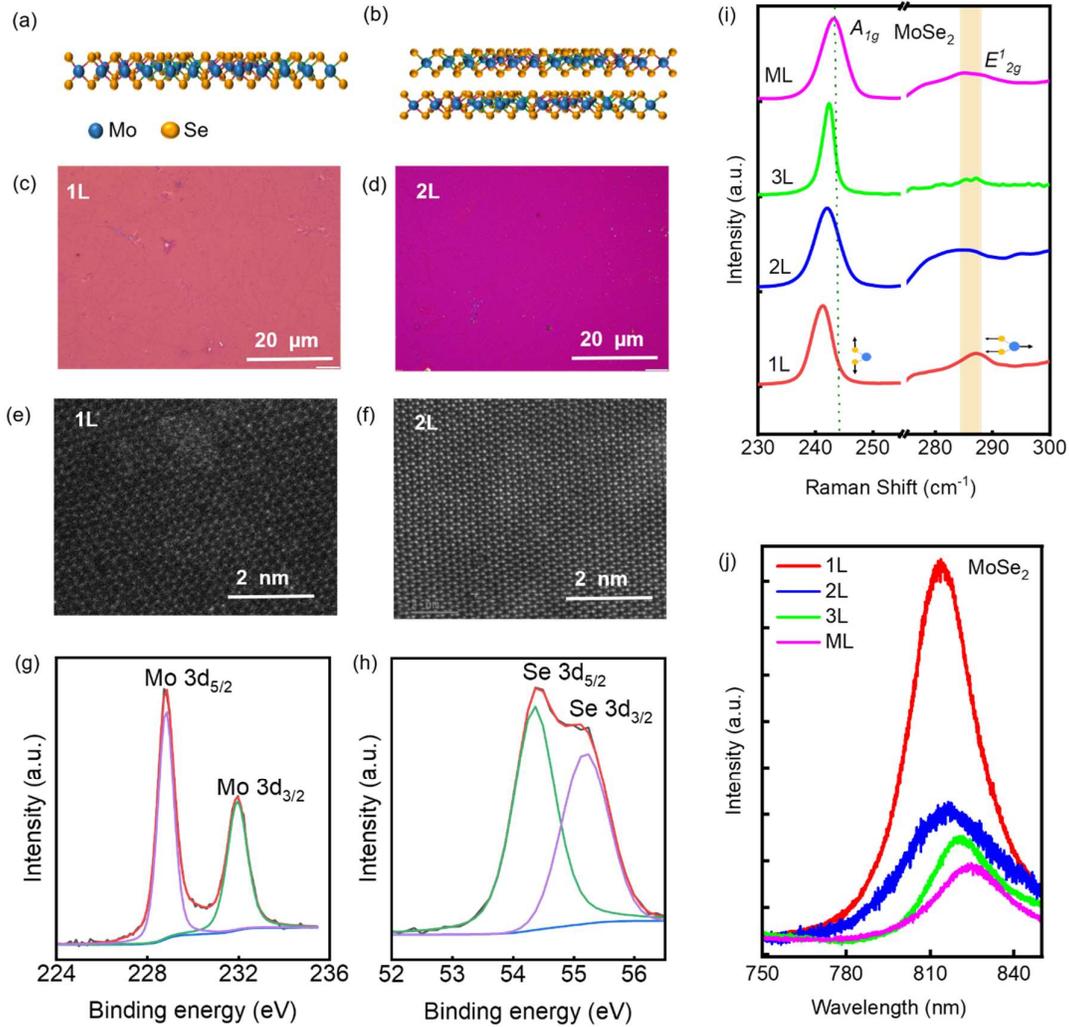

*Figure 1. (a, b) atomic model and the corresponding (c, d) optical microscopy images of the monolayer and bilayer MoSe$_2$ grown via CVD on Si/SiO$_2$ substrate, respectively; (e, f) HAADF-STEM images of monolayer and bilayer MoSe$_2$, respectively. XPS Spectra of (g) Mo 3d binding energy and (h) Se 3d binding energy. (i) the Raman and corresponding (j) PL spectra of the layer-dependent MoSe$_2$ 2D films.*

## 4.2. Electrochemical characterization

Figure 2 (a-b) displays the CV curves of the layer-dependent MoSe$_2$ 2D film at a scan rate of 50 mV/s within a potential range of 0 - 0.8 V under dark and illuminated conditions. As a semiconductor, the 2D MoSe$_2$ film absorbs light, and the applied voltage separates the electron-hole pairs generated by photoexcitation. These additional charge carriers enhance energy storage by increasing capacitance, facilitating the accumulation of significant electrolyte ions at the interface, and further boosting. Figure 2(c) illustrates the areal capacitance of 2D MoSe$_2$ films as a function of the scan rate recorded at 50 mV/s. The areal capacitance values under dark and illuminated conditions were measured as follows: 27 and 29μF/cm² for the 1L, 28 and 30μF/cm² for the 2L, 33 and 36μF/cm² for the 3L, and 57 and 75μF/cm² for the ML, respectively. Figures S6(a-b) present the CV curves of the ML- MoSe$_2$ film at varying scan rates (10, 20, 50, 80, and 100 mV/s) within a potential range of 0 - 0.8 V under both dark and illuminated conditions. Figure 2(c) demonstrates that the ML- MoSe$_2$ film achieves the maximum areal capacitance across the different scan rates, showcasing its superior energy storage capability.

Figure 2(d and e) presents the GCD curves for the layer-dependent MoSe$_2$ film at a current density of 5 μA/cm² within a potential range of 0 to 0.8 V under dark and illuminated conditions. Under illumination, photoexcitation generates electron-hole pairs in MoSe$_2$, increasing the local carrier density and enhancing charge transport. This increase in carrier population modifies the electrode/electrolyte interface, promoting greater ion adsorption and boosting charge storage capacity, as evidenced by the longer charge times. Additionally, illumination suppresses rapid carrier recombination, enabling the electrode to retain its stored charge for extended periods, as indicated by prolonged discharge times.

The areal capacitance of layer-dependent MoSe$_2$ films measured at a constant current density of 5 μA/cm² under dark and illuminated conditions is illustrated in Figure 2(f). The capacitance values observed were:

i) Monolayer: 19 μF/cm² (dark) and 20 μF/cm² (illuminated),
ii) Bilayer: 26 μF/cm² (dark) and 28 μF/cm² (illuminated),
iii) Trilayer: 40 μF/cm² (dark) and 43 μF/cm² (illuminated),
iv) Multilayer: 96 μF/cm² (dark) and 115 μF/cm² (illuminated).

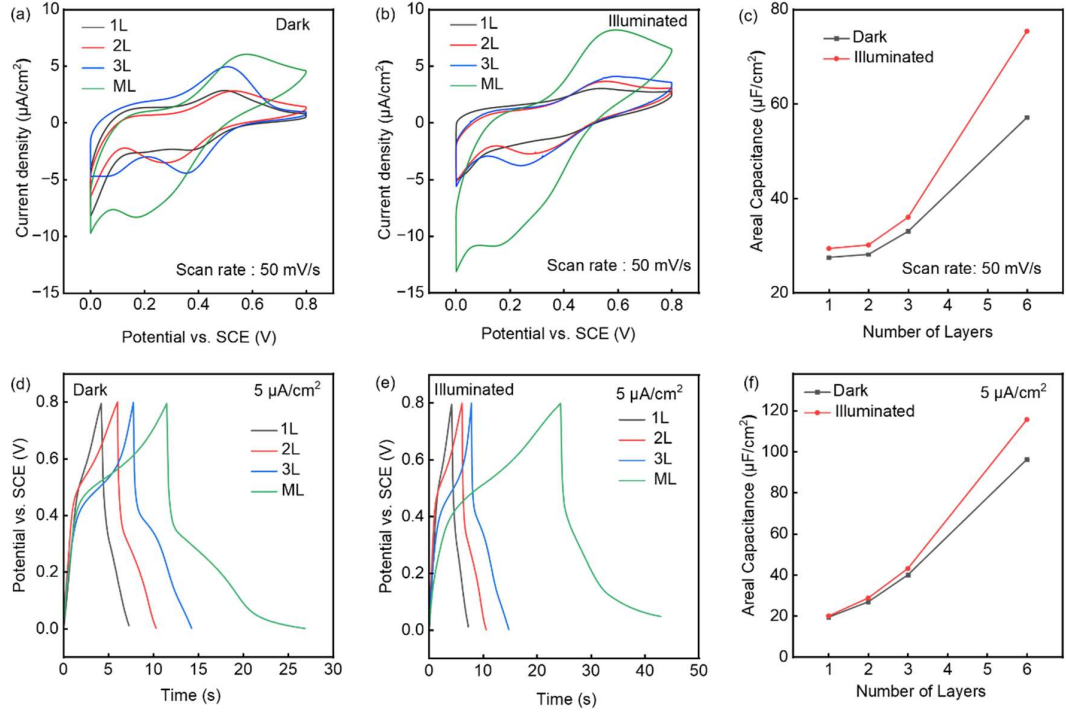

**Figure 2.** *(a-b) CV analysis of the layer-dependent MoSe$_2$ 2D films under dark and illuminated conditions at a scan rate of 50 mV/s, and the corresponding (c) Areal Capacitance variation with change in the number of 2D layers at a scan rate of 50 mV/s; (d-e) GCD analysis of the layer- dependent MoSe$_2$ thin-film under dark and illuminated conditions at a current density of 5 µA/cm$^2$, and the corresponding (f) Areal Capacitance Variation with change in the number of 2D layers at the current density of 5 µA/cm$^2$.*

Figures S6(d-e) show GCD curves of the ML-MoSe$_2$ 2D film at increasing current densities of 5, 10, and 20 µA/cm² within a potential range of 0 to 0.8 V. These curves reveal that the charge-discharge efficiency and the capacitance enhancement are more pronounced under illuminated conditions, reflecting a significant improvement in charge storage capacity with rising current density. Figure S6(f) displays the variation in the areal capacitance of the ML-MoSe$_2$ 2D film across different current densities. These results confirm that the ML-MoSe$_2$ 2D film achieves maximum areal capacitance, demonstrating superior charge storage capacity across diverse experimental conditions.

### 4.3. Charge-Storage Mechanism of the Multi-layer MoSe$_2$ film Electrode

The charge storage mechanisms of the ML-MoSe$_2$ 2D film electrode were investigated using CV measurements, analyzing the capacitive contributions in both dark and illuminated conditions. The total capacitance of the electrode materials arises from two main processes: the diffusion-controlled process and the surface-controlled capacitive process[26]. Each contributes differently to the overall capacitance under varying lighting conditions, as demonstrated through CV data.

$$(MoSe_2) + H^+ + e^- \leftrightarrow \text{MoSe} - \text{SeH}^+ \qquad \textit{(Faradaic process)} \qquad (13)$$

$$(MoSe_2)_{surface} + H^+ + e^- \leftrightarrow (MoSe_2 - H^+)_{surface} \quad \text{(Non-Faradaic process)} \quad (14)$$

To differentiate between the diffusion-controlled and capacitive contributions, the power law $I = av^b$ was applied to the CV data, where $a$ and $b$ are adjustable parameters, and $v$ is the scan rate [Figure 3(a)]. Generally, in ideal diffusion-controlled faradaic processes, the Cottrell equation predicts $b = 1/2$, whereas non-diffusion-controlled capacitive processes exhibit $b = 1$. Consequently, the total current $I$ in CV at a given potential (V) can be expressed as the sum of diffusion-controlled current ($k_2 v^{1/2}$) and capacitive currents ($k_1 v$)

$$I(V) = k_1 v + k_2 v^{1/2} \quad (15)$$

$$\frac{I(V)}{v^{1/2}} = k_1 v^{1/2} + k_2 \quad (16)$$

The capacitive and diffusion-controlled current contributions at various potentials can be distinguished by calculating coefficients $k_1$ and $k_2$ from the slope and intercept of Equation (16), respectively[51]. Subsequently, Figure 3(c-d) illustrates the segregation of capacitive and diffusion-controlled currents at different scan rates, including 10, 20, 50, 80, and 100 mV/s. Notably, the contribution from surface capacitance increases progressively with the scan rate, indicating enhanced surface-controlled charge storage at higher scan rates.

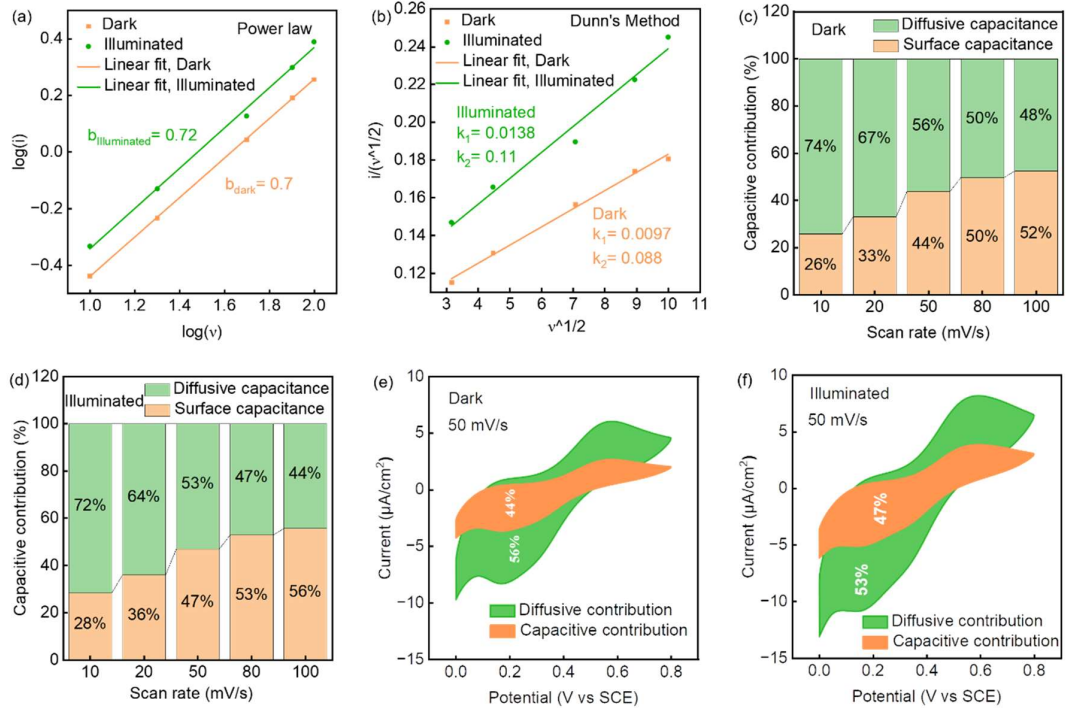

*Figure 3. (a) Logarithmic plot of current density vs. scan rate under dark and illuminated conditions, (b) The Linear plot of i/(v$^{1/2}$) vs v$^{1/2}$ under dark and illuminated conditions. (c-d) Diffusive and surface Capacitance contribution under dark and illuminated conditions at different scan rates and the corresponding (e-f) current at a scan rate of 50 mV/s.*

To investigate the charge-transfer kinetics at the interface between the electrode and electrolyte, EIS analysis was performed using layer-dependent MoSe$_2$ 2D films. Figure 4(a-c) presents the Nyquist plots of these electrodes under both dark and illuminated conditions, highlighting two distinct regions. In the high-frequency region, a semicircle is observed, which provides information about the solution resistance ($R_s$) and charge transfer resistance ($R_{ct}$). The diameter of the semicircle reflects the charge transfer resistance. In the low-frequency region, a straight line—referred to as the Warburg line—indicates Warburg impedance, representing the diffusion of electrolytic ions. The Nyquist plots of bare FTO [Figure 4(a)] and 2D MoSe$_2$ electrodes in the high-frequency region [Figure 4(b,c)], with different layer thicknesses, show variations in electrical resistance under dark and illuminated conditions. Under illumination, the $R_s$ of the 1L-MoSe$_2$ electrode decreases by 3% from ~6.1 Ω/cm², while the $R_{ct}$ reduces by 2% from 19.2 Ω/cm². For ML-MoSe$_2$, $R_s$ decreases by 12% from 8.2 Ω/cm², and $R_{ct}$ drops by 5% from 20 Ω/cm² (Table S1). These decreases indicate that photoexcitation enhances charge transfer at the electrode-electrolyte interface, improving the electrochemical performance of the electrode.

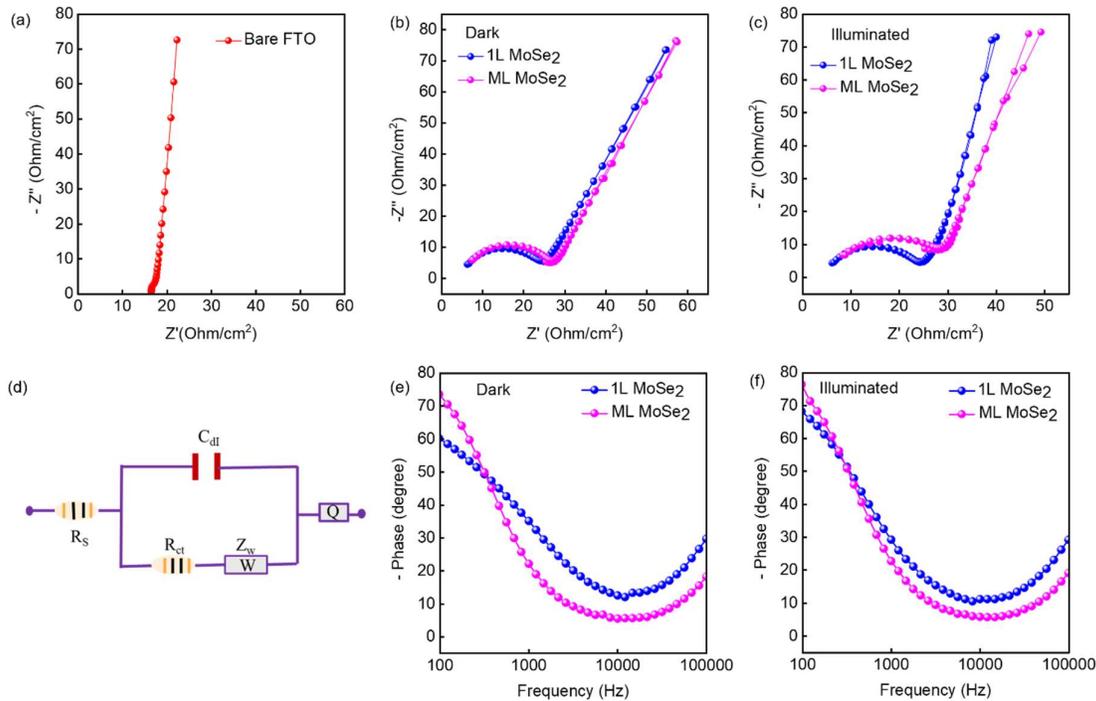

*Figure 4.* (a) Nyquist plot of bare FTO and (b-c) for 1L and ML MoSe$_2$ 2D films under dark and illuminated conditions, (d) Randles circuit diagram and (e-f) Bode plots of 1L and ML MoSe$_2$ thin-films under dark and illuminated conditions.

A typical Randles circuit diagram was used for all electrodes with $R_s$, $R_{ct}$, double layer capacitance (C$_{dl}$), constant phase element (CPE, Q), and Warburg Impedance (W) [Figure 4(d)]. The EIS fitting analysis for all Nyquist plots was carried out using ZSimpWin software. Figure 4(e,f) shows the Bode plots (phase angle *vs.* frequency) obtained from EIS for MoSe$_2$ 2D electrodes of different layer thicknesses under dark and illuminated conditions. Under illumination, the maximum phase angles are observed as 65.8° and 71.4° for 1L and ML electrodes, respectively, surpassing the corresponding values of 58.3° and 70.6° observed in the dark. This increase suggests that light enhances the capacitive behavior of the system, likely by promoting charge generation and transport. At a frequency of 120 Hz, the phase angle approaches 90°, indicating nearly ideal capacitive characteristics. Generally, phase angles closer to 90° signify strong capacitive performance, while lower phase angles suggest more resistive or diffusive processes.

## 5. Theoretical Insights into Excitonic Properties and Quasiparticle Band Structures:

In layered TMDs like MoSe$_2$, various capacitive phenomena contribute to the overall capacitance, including electric double-layer capacitance (EDLC), quantum capacitance, and diffusive capacitance. Each mechanism plays a distinct role, but quantum capacitance may become dominant, especially when exposed to light [52]. Herein, the effect of sample thickness, i.e., number of layers, and photo-induced boost in the capacitive performance were explored.

Band nesting is one prominent factor contributing to the enhanced quantum capacitance via light absorption in two-dimensional MoSe$_2$. Band nesting occurs when the conduction and valence bands are aligned parallel to each other at specific k-points, fulfilling the criteria where $|\nabla_k(E_C - E_V)| = 0$ with $|\nabla_k E_C| \approx |\nabla_k E_V| > 0$, with $E_C$ and $E_V$ denoting the corresponding conduction and valence band energies, whereas $\nabla_k$ representing the gradient of energy with respect to momentum[53]. Optical transitions at designated k-points initiate the generation of electron-hole pairs, which propagate with equal but opposite velocities in k-space. These pairs spontaneously separate and subsequently relax toward the band extrema, enhancing their effectiveness as photo-current converters[54]. As Figure 5(a) illustrates, monolayer and bilayer MoSe$_2$ exhibit band nesting—particularly along the Γ-Q and M-Γ high-symmetry vectors. In the monolayer, the strong direct-gap alignment near K underpins robust nesting and high optical absorption. However, when stacking additional layers to form a bilayer, interlayer coupling begins to split the bands at the Γ and Q valleys. A pronounced band-nesting was estimated in trilayer and 6-layered MoSe$_2$ [Figure 5(b)], where interlayer hybridization broadens the bands and diminishes the perfect parallel dispersion. Consequently, the valence band maximum shifts away from K towards Γ, while the conduction band minimum moves closer to Q. These changes disrupt the near-parallel conduction–valence band alignment characteristic observed in monolayer TMDs [55]. As a result, optical absorption does not rise linearly with adding more layers, reflecting the progressive weakening of monolayer-type nesting effects under stronger interlayer coupling. While multi-layer systems (at least up to three layers) can exhibit alternative degeneracies in some areas of the Brillouin zone, these do not fully compensate for losing the original strong nesting that dominates light absorption in monolayers.

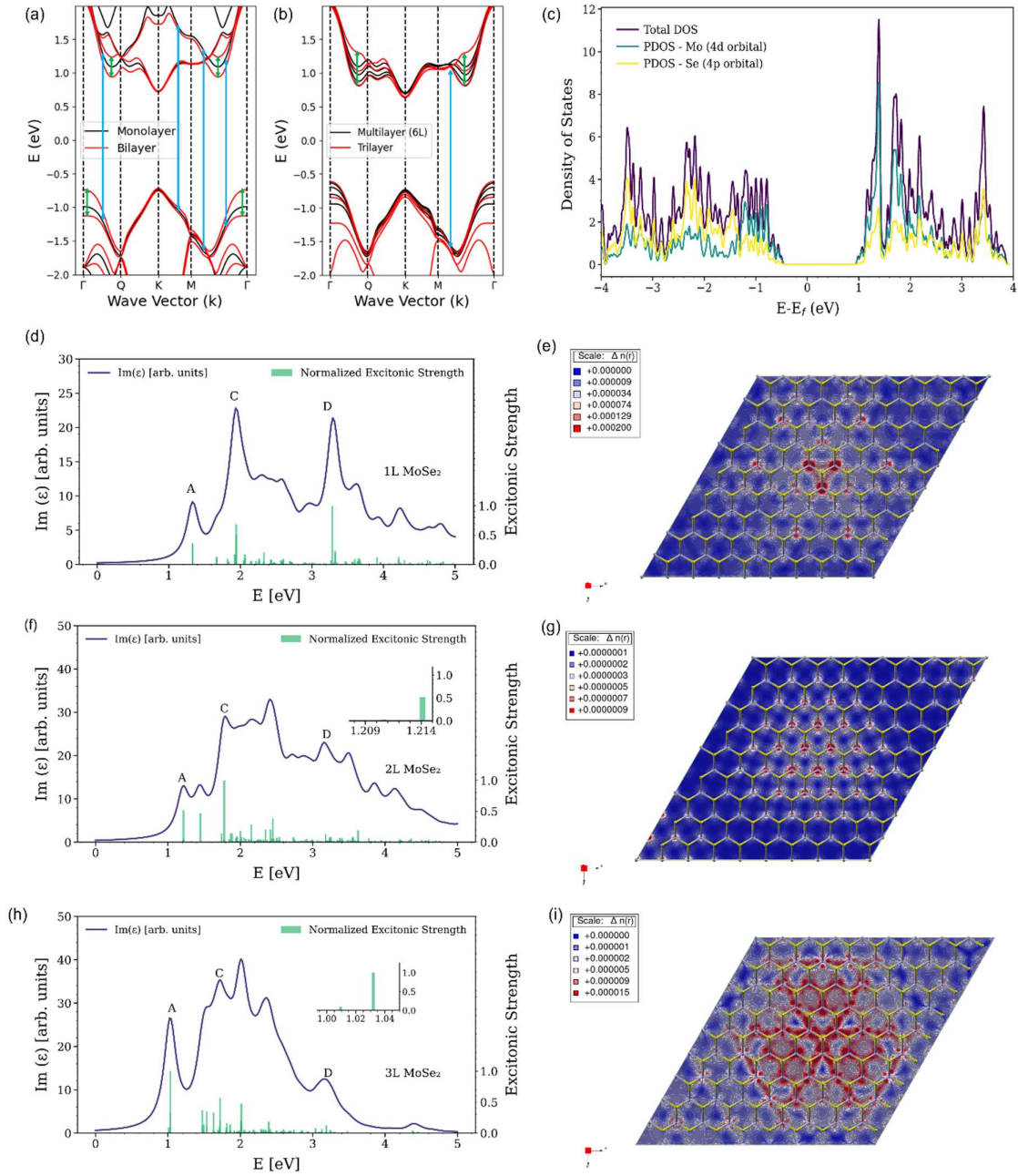

*Figure 5. Electronic band structure via KS-DFT illustrating band nesting of (a) monolayer and bilayer MoSe$_2$, (b) Trilayer and six-layer MoSe$_2$, where blue lines indicate the band nesting transitions while green lines indicate the band splitting due to added layers, (c) Density of states of monolayer MoSe$_2$ showing the van Hove singularity peaks near Fermi energy, (d, f, h) Absorption spectra and the corresponding excitonic strengths of monolayer, bilayer and trilayer MoSe$_2$ calculated with the help of Bethe-Salpeter Equation. (e, g, i) Excitonic spatial distribution of monolayer, bilayer and trilayer MoSe$_2$ where $\Delta n$ represents the electron density fluctuations.*

Figure 5(c) presents the partial and total density of states near the Fermi energy for a 2D monolayer of MoSe$_2$. The electronic DOS is primarily influenced by the 4*d* orbitals of Mo and the 4*p* orbitals of Se, as these orbitals form the bonding states within the materials. The pronounced peaks in the density of states are attributed to van Hove singularities (VHS) within the valence and conduction bands[56]. The VHS arises at specific energies where the electron group velocity approaches zero, leading to a divergence in the density of states. This effect is often due to electronic band localization or band dispersion at critical points in the Brillouin zone [57]. Mathematically, the VHS is connected to the curvature of the energy bands[58]. When the curvature becomes zero in a specific direction (as in saddle point), the density of states spike, and the electrons can occupy the states in a relatively large range of momenta near these energies[57]. This flattening of the band dispersion essentially makes the photons of corresponding energy induce transitions across broader momentum space since electrons with marginally different momenta but close enough energy levels can still satisfy the energy criteria for the transitions.[57]

In Figure 5(d), the absorption spectra and excitonic strength of monolayer $MoSe_2$ are presented, illustrating the C excitonic peak at approximately 1.95 eV. This peak is primarily attributed to optical transitions associated with band nesting in the monolayer $MoSe_2$ absorption spectra. The excitonic strength is defined as the linear combination of the square of dipole transition matrix elements between electron-hole pairs[40]. Excitonic strengths are normalized to the largest strength, resulting in the brightest exciton with the strength 1. The C exciton primarily contributes to absorption at higher energies and predominantly involves the transitions due to band nesting. The D excitonic peak at 3.3 eV predominantly arises from transitions between the K and M directions. This D exciton shows considerable passivation as the number of layers increases in the system. The excitonic strengths related to the corresponding excitonic peaks illustrate that C exciton possesses the highest strength compared to other excitonic peaks, suggesting the profound impact of band nesting in the enhanced photo-absorption observed in monolayer $MoSe_2$.

The modifications in excitonic peaks induced by adding a second layer of $MoSe_2$ are detailed in Figures 5(f) and 5(h), demonstrating how increasing the number of layers influences the optical properties of the material. In our analysis, the combined A and B excitonic peaks correspond to optical transitions at the K point in the Brillouin zone. In bilayer $MoSe_2$, we identify two distinct lowest excitonic states that arise from the symmetric and antisymmetric combinations of excitonic wave functions, a phenomenon known as Davydov splitting [59]. Typically, the energy difference between these two Davydov-split components is only a few meV, making them difficult to resolve in absorption spectra. A key feature of Davydov splitting is that one exciton is optically dark while the other is bright [60,61]. Specifically, Figure 5(f) shows that the two lowest excitonic states of bilayer $MoSe_2$ are separated by 4 meV, while Figure 5(h) demonstrates a Davydov splitting of 23 meV in the case of trilayer $MoSe_2$. The magnitude of the energy splitting indicates the extent of electronic interaction between adjacent layers [62]. An increase in splitting suggests that interlayer interactions grow stronger from bilayer to trilayer, which notably disrupts the degree of band nesting compared to the bilayer configuration.

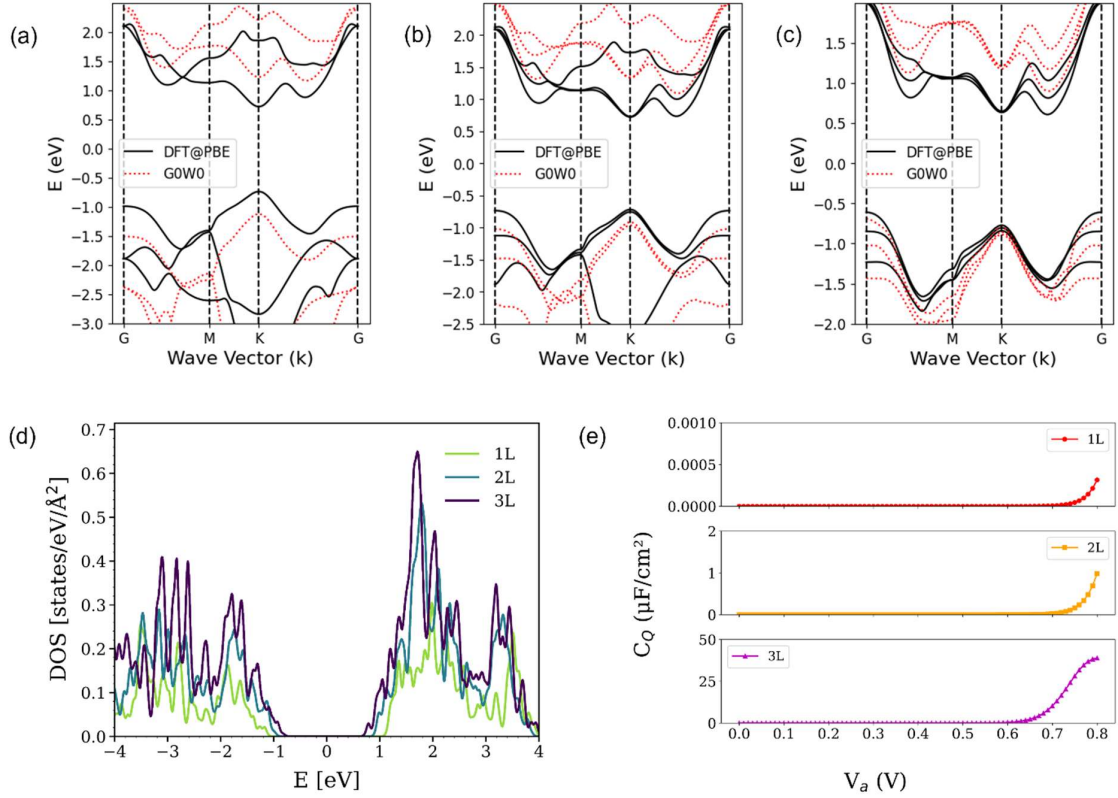

*Figure 6:* Comparison of $G_0W_0$ and DFT band structures using the PBE exchange-correlation functional for (a) monolayer [DFT band gap – 1.49 eV (direct); $G_0W_0$ band gap – 2.35 eV (indirect)], (b) bilayer [DFT band gap – 1.46 eV (direct); $G_0W_0$ band gap – 1.95 eV (indirect)], (c) trilayer [DFT band gap – 1.23 eV (indirect); $G_0W_0$ band gap – 1.6 eV (indirect)], (d) Density of states calculated via $G_0W_0$ and normalized over area, (e) Layer-dependent plot of quantum capacitance from the Density of States (DOS) calculated at the $G_0W_0$ level.

As the number of layers increases, optical absorption increases, as expected. However, in relation to the lowest excitonic peak, the C excitonic peak—arising from band nesting—shows a relatively smaller increase. In contrast, the D excitonic peak displays almost no increase from the monolayer to the bilayer but diminishes significantly between the bilayer and trilayer configurations. This further supports the notion that optical absorption is considerably affected during the transition from monolayer to multilayer, suggesting that quantum capacitance does not increase linearly with the addition of layers.

The spatial excitonic distribution for the C exciton in monolayer [Fig. 5(e)], bilayer [Fig. 5(g)] and trilayer $MoSe_2$ [Fig. 5(i)], respectively, represents the probability of finding an electron at a position ***r*** when a hole is fixed at **r'**. The position r' is chosen where the valence electrons contributing to the exciton are localized. Given the higher electronegativity of S atoms, the holes are fixed just above the S atoms to prevent them from being positioned at the center of the atom. In the monolayer case, the excitonic amplitude (red regions) is relatively compact and more strongly localized. As the number of layers increases, the exciton wavefunction extends further across the plane. This behavior aligns with the increase in dielectric screening, which reduces the exciton's binding energy and leads to more spatial delocalization.

Figure 6(a-c) illustrates the band structures of monolayer, bilayer, and trilayer $MoSe_2$ as calculated via DFT and the $G_0W_0$ approximation. The solid black lines represent the band structure calculated using DFT, while the red dotted lines indicate modifications under the state-of-the-art $G_0W_0$ approximation. These visualizations highlight the method-dependent differences in predicting electronic properties, which is substantial in this case. Notably, the $G_0W_0$ band structure shows an indirect band gap in the case of monolayer $MoSe_2$, whereas the DFT band structure illustrates the direct band gap at the K point. Similar inconsistency was reported for the sternheimerGW calculations by Zibouche *et. al.*,[63] where the reason for such direct-to-indirect nature was assigned to the dielectric screening effect of the substrate[64].

Figure 6(d) presents the DOS calculated within the $G_0W_0$ framework for monolayer, bilayer and trilayer $MoSe_2$. Notably, the density of states near the Fermi level (Eq. (5)), which plays a crucial role in influencing the quantum capacitance, exhibits a monotonic increase due to the increment in the number of electrons per unit area in the system. This accounts for the increased quantum capacitance with the number of layers. Furthermore, the precise prediction of the electronic band gap in $G_0W_0$ calculation results in realistic quantum capacitance values. Figure 7(e) presents the quantum capacitance calculated using the $G_0W_0$ approximation. The DOS calculated via $G_0W_0$ within the YABMO, is expressed in units of [states/Ha]. To correctly incorporate DOS values into the equation, we normalized the DOS by the area and converted the energy in eV, yielding quantum capacitance in units of [μF/cm²]. This allows direct comparison of theoretical and experimentally measured quantum capacitance values. Consistent with our experimental data, this figure confirms the trend of increasing capacitance with added layers and higher gate voltages. The maximum quantum capacitance was observed at 0.8 eV for all the samples with 4 nF/cm² for monolayer, 1.1 μF/cm² for bilayer, and 38 μF/cm² for trilayer $MoSe_2$, respectively.

## 6. CONCLUSION

This study successfully synthesized 2D $MoSe_2$ (2H phase) films with controlled layer thicknesses using the APCVD method. The findings demonstrate a layer-dependent increase in areal capacitance and further enhancement under light illumination. Notably, multilayer $MoSe_2$ films exhibited a significant increase in areal capacitance upon illumination, rising from 96 μF/cm² in dark conditions to 115 μF/cm² at a current density of 5 μA/cm². Photon absorption generates electron-hole pairs in TMDs, increasing the DOS at the Fermi level and enhancing quantum capacitance without significantly affecting EDLC or diffusive capacitance. This highlights the potential of utilizing light-responsive $MoSe_2$ for optoelectronic and solar-powered applications. Theoretical investigations using DFT and MBPT confirm that band nesting and Van Hove singularities are crucial in enhancing optical absorption and quantum capacitance. However, GW and BSE calculations were performed up to trilayer MoSe$_2$ due to computational constraints. A dedicated study focusing on 6-layered MoSe$_2$ is necessary to fully capture the excitonic and carrier dynamics in large systems. While the precise theoretical framework for light-induced quantum capacitance modulation remains an open area of research, our study demonstrates that increased charge carriers in the conduction band enhance quantum capacitance, leading to improved capacitive performance under illumination. Furthermore, ab initio calculations align with experimental trends, confirming that capacitive performance strengthens with additional layers.

Overall, these findings emphasize the effectiveness of APCVD in fabricating high-quality 2D MoSe$_2$ films and their potential for smart energy storage and optoelectronic applications with light-tunable performance. Future studies should focus on device integration, such as hybrid supercapacitors and photodetectors, while advancements in large-area synthesis could enhance scalability for flexible electronics, self-powered sensors, and energy harvesting technologies.

# SUPPORTING INFORMATION

**Table S1**: The Solution Resistance and Charge Resistance Values from Nyquist plots.

| Electrodes | $R_S$ (Ω/cm$^2$) in dark | $\Delta R_S$ (Ω/cm$^2$) in Light (%) | $R_{ct}$ (Ω/cm$^2$) in dark | $\Delta R_{ct}$ (Ω/cm$^2$) in Light (%) |
|---|---|---|---|---|
| 1L MoSe$_2$ | 6.1 | 3 | 19.2 | 2 |
| ML MoSe$_2$ | 8.2 | 12 | 20 | 5 |

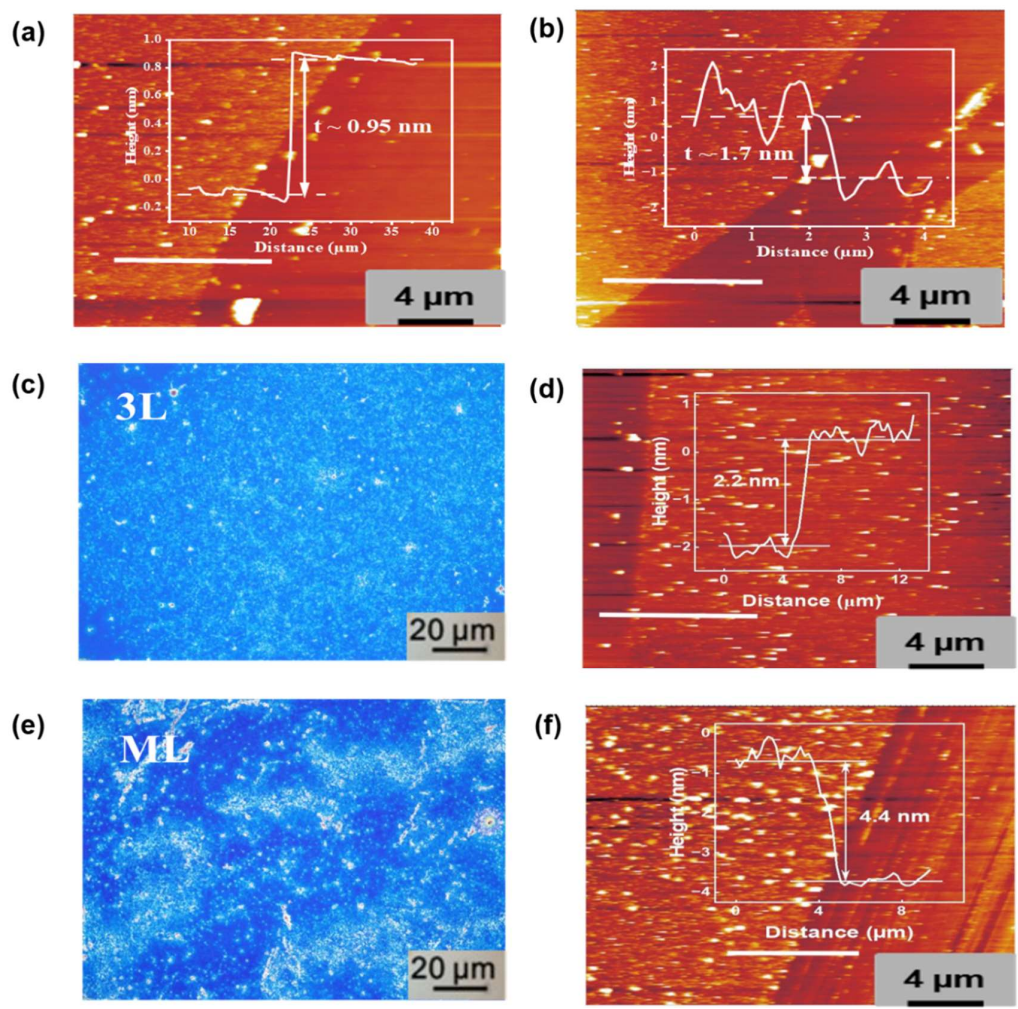

**Figure S1: (a, b)** the AFM image along with the height profile of the monolayer and bilayer MoSe$_2$, **(c, f)** the optical microscopy image and AFM image along with the height profile of the trilayer and multilayer MoSe$_2$.

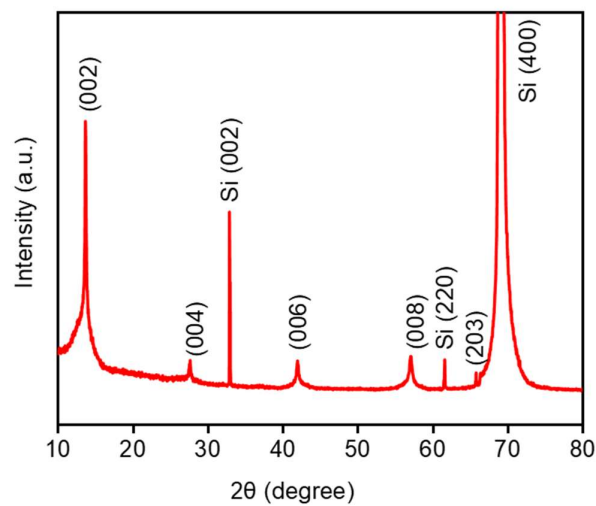

**Figure S2:** XRD characterizations of the multilayer (6 layers) MoSe$_2$ thin-film indicating the high crystalline quality of the as-grown 2D film.

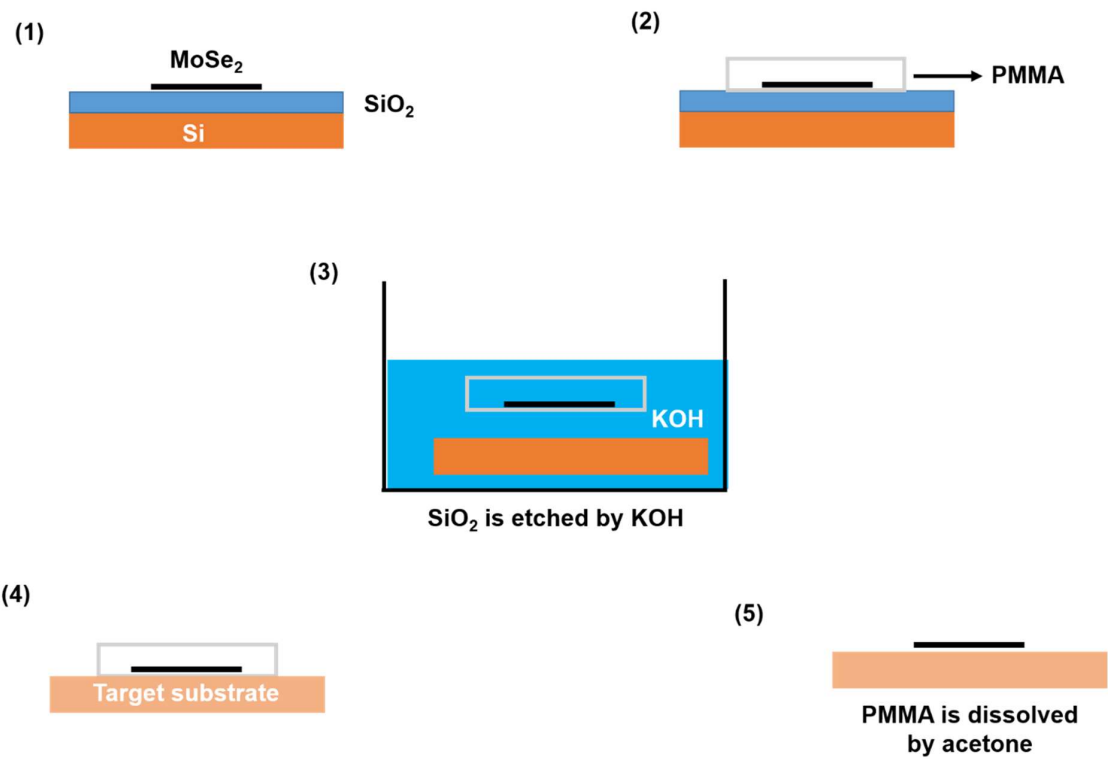

**Figure S3:** Schematic representation showing the polymer-assisted wet transfer process for MoSe$_2$ thin-film from as-synthesized SiO$_2$ substrate to FTO substrate.

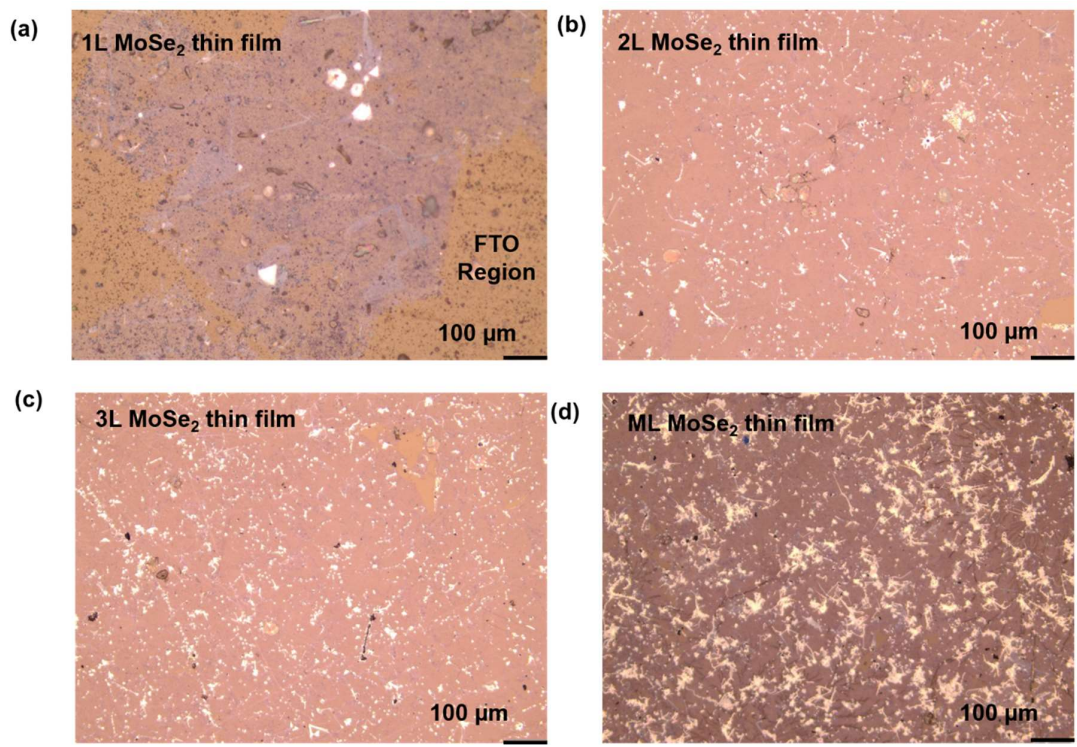

**Figure S4:** Optical images of Monolayer, Bilayer, Trilayer and Multilayer MoSe$_2$ thin films on FTO Substrate after transfer with PMMA-assisted method.

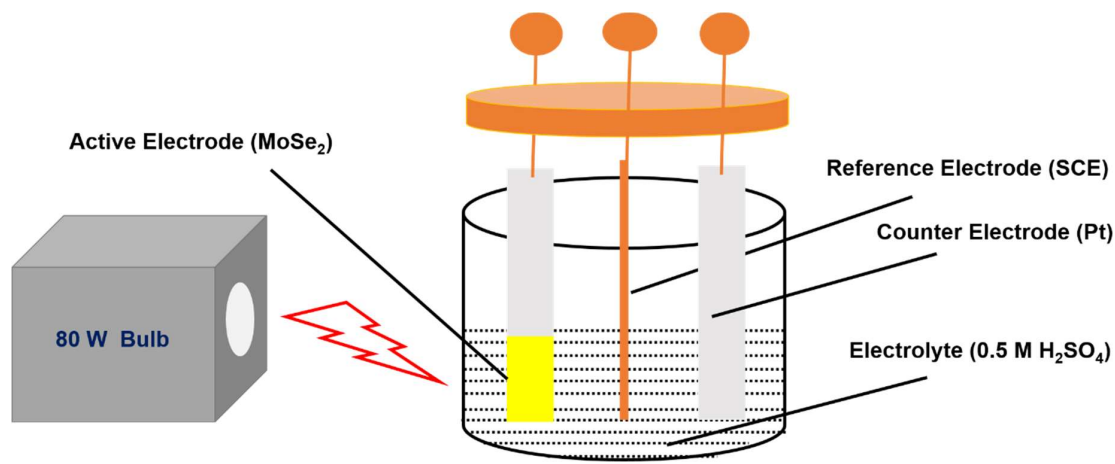

**Figure S5:** Schematic of three-electrode configuration in electrochemical measurements used for this study.

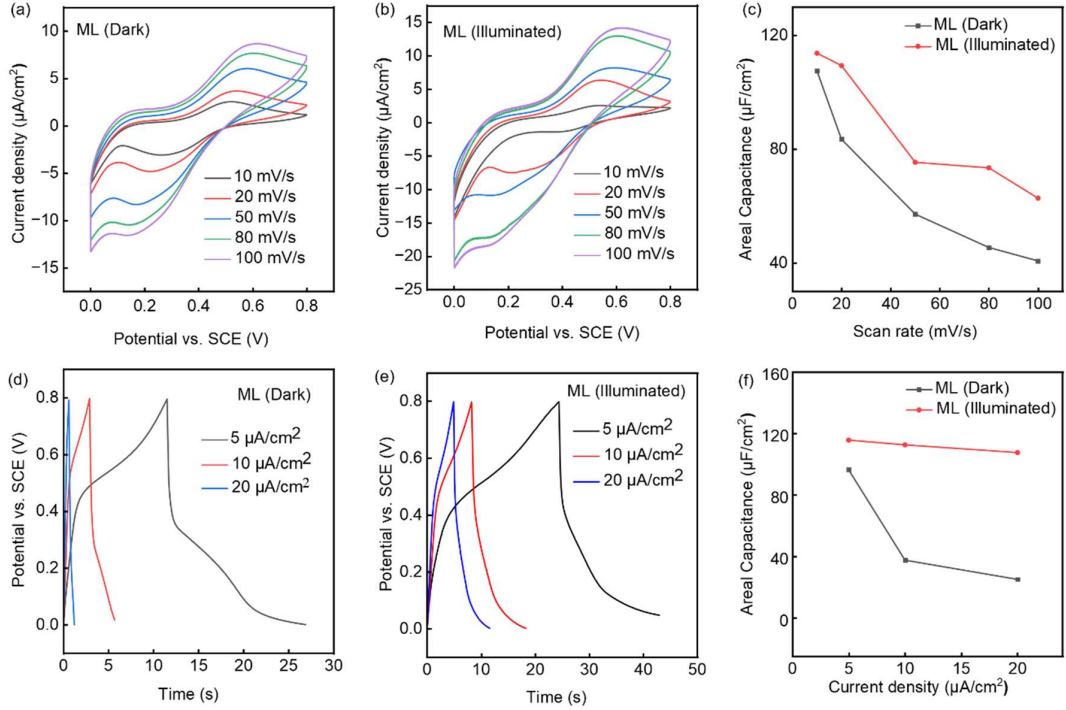

**Figure S6:** (a,b) CV analysis of the multilayer MoSe$_2$ thin-film under dark and illuminated conditions at different scan rates and the corresponding (c) Areal Capacitance. (d,e) GCD analysis of the multilayer MoSe$_2$ thin-film under dark and illuminated conditions at different current densities and the corresponding (f) Areal Capacitance.

**Effect of the Density of States near the Fermi Energy Level on Quantum Capacitance:**

The quantum capacitance for 2D material is given by,

$$C_Q = e^2 \int_{-\infty}^{+\infty} \frac{D(E)}{4K_BT} \text{sech}^2 \left(\frac{E - eV_a}{2K_BT}\right) dE$$

As 'E' is a dummy variable, we can modify the above equation as,

$$C_Q = e^2 \int_{-\infty}^{+\infty} \frac{D(E - E_f^{V=0})}{4K_BT} \text{sech}^2 \left(\frac{E - E_f^{V=0} - eV_a}{2K_BT}\right) dE,$$

$$C_Q = e^2 \int_{-\infty}^{+\infty} \frac{D(E - E_f^{V=0})}{4K_BT} \text{sech}^2 \left(\frac{E - E_f^{V=V_a}}{2K_BT}\right) dE$$

We can write the equation in summation form of discrete energies and their corresponding Density of States calculated in DFT (or) MBPT calculation,

$$C_Q = \frac{e^2}{4K_BT} \sum_i D(E_i - E_f^{V=0}) \text{sech}^2\left(\frac{E_i - E_f^{V=V_a}}{2K_BT}\right)$$

Here, $\text{sech}^2\left(\frac{E_i - E_f^{V=V_a}}{2K_BT}\right)$ serves as a weighting factor for $D(E_i - E_f^{V=0})$ in the summation.

As indicated in Fig. 1, these weights are significant only near the Fermi energy. This concludes that only the states close to the Fermi energy will contribute significantly to the quantum capacitance."

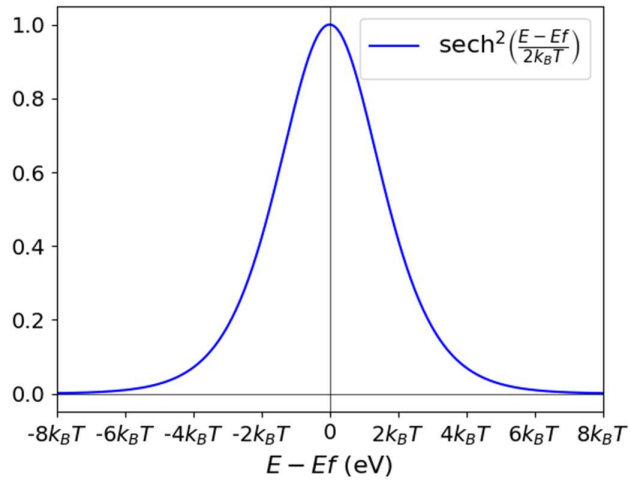

**Figure S7:** Plot of $\text{sech}^2\left(\frac{E-E_f}{2K_BT}\right)$ term in the equation of Q$_C$, which acts as weights for the corresponding density of states.

**Convergence Criteria:**

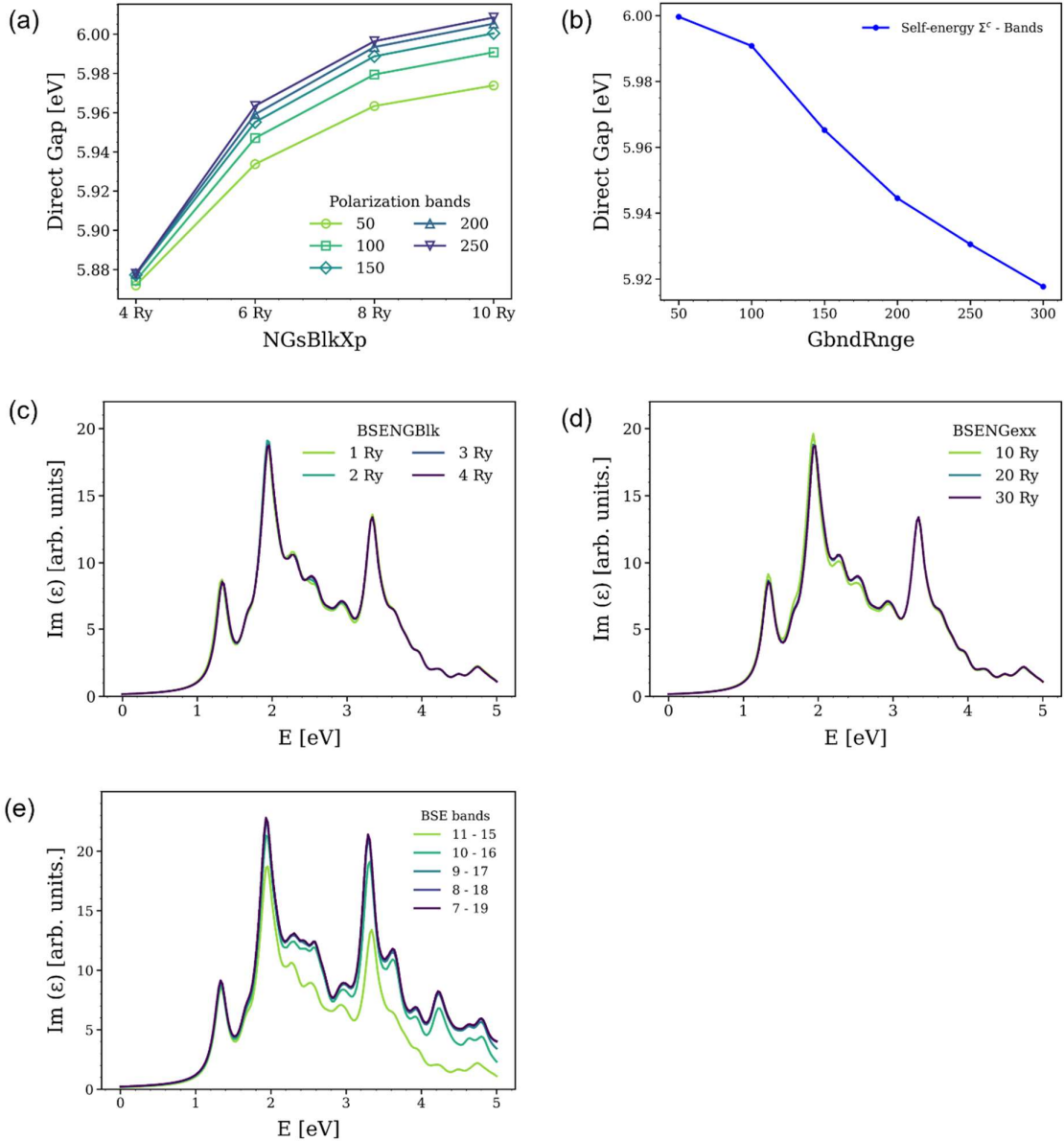

**Figure S8:** Convergence tests for key computational parameters in the many-body perturbation theory framework: **(a)** Polarization function convergence with respect to the number of G-vectors (NGsBlkXp) and bands included in the response function, **(b)** Correlation self-energy convergence evaluated by varying the number of bands (GbndRnge) in the correlation kernel. **(c)** Exchange contribution to the BSE kernel: convergence of summed G-components (BSENGexx), **(d)** Screened Coulomb potential matrix $W(G, G')$: convergence tests for reciprocal lattice (RL) vector components (BSENGBlk) in the electron-hole (e-h) attractive kernel, **(e)** E-h basis size dependence: convergence of the BSE kernel with respect to the number of bands in the e-h excitation subspace.